\documentclass[journal]{IEEEtran}

\usepackage[utf8]{inputenc}
\usepackage[T1]{fontenc}
\usepackage{cite}
\usepackage[cmex10]{amsmath}
\usepackage{amssymb}
\usepackage{mathtools}
\usepackage{url}
\usepackage{xcolor}
\usepackage{needspace}
\usepackage[hidelinks]{hyperref}

\DeclareMathOperator*{\argmin}{arg\,min}

\DeclareMathOperator{\prox}{prox}
\newcommand{\R}{\mathbb{R}}
\newcommand{\E}{\mathbb{E}}
\newcommand{\norm}[1]{\left\lVert #1 \right\rVert}

\newtheorem{theorem}{Theorem}
\newtheorem{lemma}{Lemma}

\newtheorem{assumption}{Assumption}

\begin{document}

\title{Deriving Approximate Message Passing
       from the Convex Gaussian Min--Max Theorem}

\author{Vikrant~Malik and Babak~Hassibi\thanks{Vikrant Malik and Babak Hassibi are with the Department of
    Electrical Engineering, California Institute of Technology (Caltech),
    Pasadena, CA, USA (e-mail: vmalik@caltech.edu;
    hassibi@caltech.edu).}}

\maketitle

\begin{abstract}
Approximate message passing (AMP) provides fast iterative algorithms
for high-dimensional signal recovery with Gaussian design matrices,
while the Convex Gaussian Min--max Theorem (CGMT) gives a static
optimization framework for obtaining sharp asymptotic characterizations
of convex estimators.  Although these two frameworks often lead to the
same scalar state-evolution equations, their connection is usually
indirect.  In this paper, we establish a direct connection between the two for regularized
linear regression in the proportional high-dimensional regime.
When the CGMT Auxiliary Optimization (AO) and Primary
Optimization (PO) give the same primal-dual solution, we show that the
CGMT framework recovers the AMP fixed-point equations, including the
Onsager correction.
We further identify the AO Gaussian vectors with
the Gaussian perturbations in the primal and residual AMP channels. For regularized $M$-estimation, the same viewpoint recovers the fixed point of scalar-variance
max-sum Generalized AMP (GAMP).
These results show that the AMP (and GAMP) fixed-point equations can be recovered from the CGMT framework and may suggest AMP-like algorithms in settings where standard derivations are unavailable.
\end{abstract}


\section{Introduction}
Signal recovery from noisy linear measurements is a central problem in
high-dimensional statistics, information theory, and machine learning.
Given observations of the form $y=Ax_0+w$, with
$A\in\R^{m\times n}$, $x_0\in\R^n$, and $y,w\in\R^m$, the goal is to
reconstruct the unknown signal $x_0$ from $m$ measurements in ambient
dimension $n$. This model captures a wide range of
applications, including compressed
sensing, sparse regression, imaging, wireless communication, and
feature selection in modern data analysis.

Over the past few years, substantial effort has been devoted to
understanding the performance characterizations for standard
 high-dimensional estimators (linear regression and its regularized
versions) in the proportional high-dimensional regime, where the
ambient dimension and number of measurements grow at a
fixed ratio. A recurring theme in the literature is to analyze the
resulting random optimization problems via either the CGMT (a convex extension of Gordon's Gaussian
min-max comparison inequality
\cite{gordon1985inequalities,stojnic2013framework,
thrampoulidis2015regularized,thrampoulidis2018mestimators}), or via
AMP
\cite{DMM09,BM11,javanmard2013stateevolution,rangan2011generalized}.
When the measurement matrix is Gaussian, both approaches
show that the estimation error is described by just a few
scalar quantities which can be obtained by solving a deterministic,
scalar optimization problem. These same quantities can then be used to
predict performance metrics like the norm and geometry of the
recovered signal, margins, and the resulting estimation
or classification error \cite{thrampoulidis2018mestimators,BM11}. In
\cite{thrampoulidis2018mestimators}, a general CGMT-based framework is
developed that yields asymptotically exact predictions for broad
classes of losses and regularizers, and characterizes performance in
the proportional regime where $m$ and $n$ grow proportionally. In the
supervised learning literature, explicit test-error characterizations
have been derived for linear models learned via
regression-style estimators under Gaussian-mixture-type
models \cite{akhtiamov2023regularized}. Another closely related work
\cite{taheri2020sharp} provides sharp benchmarks for parameter
estimation with binary observations.
More broadly, the CGMT has been used to analyze nonlinear Gaussian
models \cite{thrampoulidis2019lifting}, binary and Gaussian-mixture
classification \cite{taheri2020optimality,kini2020analytic,
deng2019double}, ridge-regularized ERM in generalized linear models
\cite{taheri2021fundamental}, multiclass classification
\cite{thrampoulidis2020multiclass}, and distributionally robust binary
learning \cite{malik2026dro}.

Despite their common asymptotic predictions, CGMT and AMP have usually
been viewed as tools of a different nature.  The CGMT is a static
comparison theorem.  It compares a
Primary Optimization \eqref{eq:po} (PO) with an Auxiliary Optimization \eqref{eq:aux} (AO):
\begin{align}
\Phi(A)&=\min_{w\in\mathcal{S}_w}\max_{u\in\mathcal{S}_u}\;
u^\top A w+\psi(w,u), \label{eq:po}\\
\phi(g,h)&=\min_{w\in\mathcal{S}_w}\max_{u\in\mathcal{S}_u}\;
\norm{w}_2\, g^\top u+\norm{u}_2\, h^\top w+\psi(w,u),
\label{eq:aux}
\end{align}
Here $w\in\R^n$ and $u\in\R^m$ are optimization variables; $A,g,h$ have independent standard
normal entries, and $\psi$ is convex--concave.
Under convexity
and compactness assumptions, the optimal values of \eqref{eq:po} and \eqref{eq:aux} concentrate together,
and properties of optimizers can be transferred from the AO to the PO.
This is stated formally later in Theorem~\ref{thm:cgmt}.  
The analytic simplification in the AO comes from the
replacement of the matrix bilinear term $u^\top A w$ by the two
decoupled Gaussian terms $\norm{w}_2 g^\top u$ and
$\norm{u}_2 h^\top w$.  This decoupling makes the AO amenable to
analysis.  It also motivates the question pursued in this paper: for a
given PO solution, which auxiliary Gaussian vectors $g,h$ would make
the AO KKT system select the same primal-dual solution?

Within this framework, we first study the quadratic regularized
regression problem
\begin{equation}\label{eq:rlr}
  \hat{x}\in\argmin_{x\in\R^n}
  \left\{\frac12\norm{y-Ax}_2^2+\lambda F(x)\right\},
\end{equation}
where $F$ is a separable convex regularizer and $\lambda>0$.  Setting
$e=x-x_0$ and $G=\sqrt m\,A$, the Fenchel dual of the
quadratic loss gives the min--max
Primary Optimization
\begin{equation}\label{eq:po_rlr}
  \begin{aligned}
  \Phi(G)&=\min_{e\in\R^n}\max_{u\in\R^m}
  \Bigl\{-\frac{1}{\sqrt{m}}u^\top G e+u^\top w\\
  &\hspace{2.4em}
  -\frac12\norm{u}_2^2+\lambda F(x_0+e)\Bigr\}.
  \end{aligned}
\end{equation}
Since $G$ has i.i.d.\ standard normal entries, \eqref{eq:po_rlr} is an
instance of the CGMT PO form \eqref{eq:po}, with $e$ as the primal
variable and $u$ as the dual variable.  The corresponding AO
\eqref{eq:ao_rlr} replaces the matrix bilinear term in
\eqref{eq:po_rlr} by separated Gaussian terms, producing an
optimization with the same saddle variables but a more tractable
Gaussian structure.
To solve the quadratic objective in \eqref{eq:rlr}, the basic proximal-gradient method is
ISTA.  With stepsize $\gamma>0$, its recursions are
\begin{align}
  r^t &= y-Ax^t, \label{eq:ista_r_intro}\\
  v^t &= x^t+\gamma A^\top r^t,
  \label{eq:ista_v_intro}\\
  x^{t+1} &= \eta(v^t;\lambda\gamma),
  \label{eq:ista_x_intro}
\end{align}
The scalar map $\eta$ used here and throughout is defined by writing
$F(x)=\sum_{i=1}^n f(x_i)$ and setting
\begin{equation}\label{eq:prox}
  \begin{aligned}
  \eta(v;\theta)
  &=\prox_{\theta f}(v)\\
  &:=\argmin_{x\in\R}
  \left\{\frac12(x-v)^2+\theta f(x)\right\},
  \quad \theta>0 .
  \end{aligned}
\end{equation}
We apply $\eta$ coordinatewise to vectors.
These iterations represent proximal gradient descent applied to
\eqref{eq:rlr}.

AMP, on the other hand, is an algorithmic method whose iterates are
tracked by state evolution.  It was introduced for compressed sensing
\cite{DMM09}, formalized in \cite{BM11}, and used to characterize the
LASSO risk for Gaussian designs \cite{bayati2012lasso}.  Subsequent
extensions include broader AMP recursions, spatial coupling,
universality, nonseparable nonlinearities, and variants such as GAMP,
D-AMP, VAMP, and OAMP
\cite{javanmard2013stateevolution,donoho2013information,
bayati2015universality,berthier2020state,rangan2011generalized,
metzler2016denoising,rangan2019vamp,ma2017orthogonal}. For the linear model above, the standard AMP recursion with
$\delta_n:=m/n\to\delta$ is
\begin{align}
  v^t &= x^t+A^\top z^t,\label{eq:ampv1}\\
  x^{t+1} &= \eta_t(v^t),\label{eq:ampx1}\\
  z^t &= y-Ax^t+
  \underbrace{\frac{1}{\delta_n}z^{t-1}
  \bigl\langle\eta_{t-1}'(v^{t-1})\bigr\rangle}
  _{\text{Onsager correction}},\label{eq:ampz1}
\end{align}
where the standard initialization is $x^0=0$, $z^{0}=y$ for a
deterministic threshold sequence $(\theta_t)_{t\ge0}$.  Here,
$\eta_t$ is a coordinatewise denoising function and
$\langle\cdot\rangle$ denotes the empirical average.  For large
Gaussian design matrices, AMP is often computationally faster
compared to generic first-order convex solvers such as ISTA and
FISTA \cite{DDD04,BeckTeboulle09}, because its iterates are tailored
to the random matrix geometry through the Onsager correction term.
Thus AMP has the same proximal update structure as ISTA, but replaces
the raw residual by the Onsager-corrected residual in \eqref{eq:ampz1}.
The motivation for the correction term is that it cancels the correlation between the current
iterate and the measurement matrix $A$, making the effective input to
the denoiser behave asymptotically as the true signal plus independent
Gaussian noise.

Previous works have noticed that the state equations
arising from AMP and from CGMT reductions can be equivalent after
suitable changes of variables
\cite{deng2019double,aubin2020perceptrons,luo2025equivalence}. In this paper, we make this connection explicit for $M$-estimation. Starting from the CGMT framework, we show how
the scalar saddle point of the AO, its stationarity conditions, and the auxiliary
Gaussian vectors recover the AMP and GAMP fixed-point structures
and their corresponding stationary scalar equations.

\subsection{Contributions}
\begin{enumerate}
  \item Under the proportional high-dimensional regime with Gaussian
  design matrices, and for the regularized linear regression problem,
  we show that the fixed-point equations of the AMP recursion
  \eqref{eq:ampv1}--\eqref{eq:ampz1} can be recovered using the
  CGMT framework under the assumption that the AO and PO solutions match.  In particular, the scalar saddle point obtained
  from the CGMT auxiliary optimization determines the Onsager coefficient appearing in the fixed-point
  equations. This is formally stated in Theorem \ref{thm:main_draft}.

  \item More generally, for separable smooth output losses, we show that the same
  stationarity comparison recovers the fixed point of scalar-variance
  max-sum GAMP.  This connection is detailed in Section~\ref{sec:general_extension} and Appendix~\ref{app:general_loss}.
  \item We provide an asymptotic interpretation of the AO Gaussian vectors.
  Under the assumption that the AO and PO solutions match, $h$ determines the noise in primal AMP channel
  $v^* \approx x_0+\tau_*h$, while $g$ determines the noise in the residual AMP channel
  $z^* \approx w+\sigma_*g$.  This is formally stated in
  Theorem \ref{thm:vector_gaussian_draft} and Theorem \ref{thm:gamp_channel_extension}.

\end{enumerate}
  Potential applications include square-root LASSO
  \cite{thrampoulidis2018mestimators}. For operator-norm robust LASSO,
  this approach yields an AMP-style algorithm \cite{malik2026robustlasso}.
  For non-Gaussian designs, the CGMT AO comparison and Gaussian-channel
  interpretation require suitable universality results
  \cite{bayati2015universality}.

\subsection{Organization}
The paper is organized as follows.  Section~\ref{sec:setup} introduces
the notation, the observation model, the regularized
linear regression problem, and reviews the AMP and CGMT frameworks.
Section~\ref{sec:main} and Section~\ref{sec:general_extension} state
the main results: the recovery of the AMP and GAMP fixed-point
equations from the CGMT and an interpretation of the AO Gaussian
vectors.

\section{Notation and Problem Setup}
\label{sec:setup}

\subsection{Notation}
For a vector $a\in\R^N$, write
$\langle a\rangle=N^{-1}\sum_{i=1}^N a_i$.  We write
$X\sim p_X$ to denote that $X$ has law $p_X$, and reserve $Z$ for a
standard normal random variable independent of all other scalar
variables unless stated otherwise. The symbols
$\xrightarrow{p}$ and $\xrightarrow{\mathrm{a.s.}}$ denote
convergence in probability and almost sure convergence, respectively.
The regularizer is separable.  Throughout, $f:\R\to(-\infty,\infty]$
is proper, closed, and convex, and for $x \in \R^n$, 
$F(x)=\sum_{i=1}^n f(x_i)$.
The scalar map $\eta$ is defined in \eqref{eq:prox} and is applied
coordinatewise to vectors.
The LASSO corresponds to the choice $f(x)=|x|$.
\subsection{Problem Setup}
We consider instances of $y=Ax_0+w$ in the proportional regime
$m,n\to\infty$ with $\delta_n\to\delta\in(0,\infty)$.  The matrix
$A\in\R^{m\times n}$ has independent entries
$A_{ij}\sim\mathcal{N}(0,1/m)$ and is independent of the signal and
noise, whose entries satisfy
$x_{0,i}\stackrel{\mathrm{i.i.d.}}{\sim}p_{X_0}$ and
$w_j\stackrel{\mathrm{i.i.d.}}{\sim}p_W$.
We assume $\E[X_0^2]<\infty$ and $\E[W^2]<\infty$ and write
$\sigma_0^2:=\delta^{-1}\E[X_0^2]$ and
$\sigma_w^2:=\E[W^2]$, and study the estimator \eqref{eq:rlr}.  For \(H\sim\mathcal N(0,1)\) independent of \(X_0\), define
\(f'_+(v):=\sup_{s\in\partial f(v)}|s|\) and assume $\mathbb E\!\left[f'_+(X_0+cH)^2\right]<\infty
\quad\text{for all }c\in\mathbb R$.

\subsection{AMP analysis}
We use the AMP recursion \eqref{eq:ampv1}--\eqref{eq:ampz1}
introduced in the Introduction.  In the proximal setting considered
here, the coordinatewise denoiser is
$\eta_t(v)=\eta(v;\theta_t)$, where $\eta$ is defined in
\eqref{eq:prox}.  The associated state evolution is formally stated as
follows.

\begin{theorem}[AMP state evolution \cite{BM11}]\label{thm:amp_se}
Fix $k\ge2$ and assume that $p_{X_0}$ and $p_W$ have finite
$(2k-2)$-th moments.  Under the above Gaussian design and iid signal
and noise model, for every fixed $t\ge0$ and every pseudo-Lipschitz
$\psi_1,\psi_2:\R^2\to\R$ of order $k$,
\[
  \frac{1}{n}\sum_{i=1}^n \psi_1(x_i^{t+1},x_{0,i})
  \xrightarrow{\mathrm{a.s.}}
  \E\!\left[
  \psi_1\bigl(\eta(X_0+\tau_t Z;\theta_t),X_0\bigr)\right],
\]
\[
  \frac{1}{m}\sum_{i=1}^m \psi_2(z_i^t,w_i)
  \xrightarrow{\mathrm{a.s.}}
  \E\!\left[\psi_2(W+\sigma_t Z,W)\right],
\]
where $X_0\sim p_{X_0}$, $W\sim p_W$, and
$Z\sim\mathcal{N}(0,1)$ is independent of both.  The scalars are
defined by
\begin{equation}\label{eq:amp_se}
  \begin{aligned}
  \tau_0^2&=\sigma_w^2+\sigma_0^2,\\
  \tau_{t+1}^2&=\sigma_w^2+\frac{1}{\delta}
  \E\!\left[
  \bigl(\eta(X_0+\tau_t Z;\theta_t)-X_0\bigr)^2\right],
  \end{aligned}
\end{equation}
with $\sigma_t^2=\tau_t^2-\sigma_w^2$. \hfill$\square$
\end{theorem}

Informally, Theorem~\ref{thm:amp_se} says that, for every fixed
iteration $t$, the high-dimensional AMP iterates decouple into scalar
Gaussian channels as $m,n\to\infty$.  The input to the denoiser behaves
coordinatewise as
\[
  (v^t)_i \;\approx\; X_0+\tau_t Z,
\]
so the next estimate behaves as
$\eta(X_0+\tau_t Z;\theta_t)$.  Similarly, the residual behaves in
empirical distribution as
\[
  (z^t)_i \;\approx\; W+\sigma_t Z.
\]
Thus the full AMP trajectory is asymptotically summarized by the scalar
state variables $(\tau_t,\sigma_t)$.

\subsection{CGMT Analysis}
This subsection revisits the CGMT reduction for regularized linear
regression from \cite{thrampoulidis2018mestimators} and the scalar
saddle equations used later. The CGMT is formally stated as follows.

\begin{theorem}[CGMT \cite{thrampoulidis2018mestimators}]\label{thm:cgmt}
Let $S_w, S_u$ be convex compact sets, $\psi$ be continuous and convex-concave on
$S_w \times S_u$, and $A$, $g$ and $h$ all have entries i.i.d.\ standard normal.
Let $S$ be an arbitrary open subset of $S_w$ and $S^c := S_w \setminus S$.
Denote by $\Phi_{S^c}(A)$ and $\phi_{S^c}(g,h)$ the optimal costs of
the PO \eqref{eq:po} and AO \eqref{eq:aux}, respectively, when $w$ is
minimized over $S^c$.
Let $w_\Phi(A)$ denote a minimizer of the PO \eqref{eq:po}.
If there exist constants $\bar{\phi} < \bar{\phi}_{S^c}$ such that
$\phi(g,h) \xrightarrow{p} \bar{\phi}$, and $\phi_{S^c}(g,h) \xrightarrow{p} \bar{\phi}_{S^c}$
(\emph{convergence in probability}), then
\[
\lim_{n\to\infty} \mathbb{P}\!\bigl(w_{\Phi}(A) \in S\bigr) = 1 .
\]
\hfill$\square$
\end{theorem}

For the regularized regression problem \eqref{eq:rlr}, we use the PO
\eqref{eq:po_rlr} introduced in the Introduction, with
$e=x-x_0$ and $G=\sqrt m A$.  We now record only the corresponding AO
and its scalar reduction.
The corresponding auxiliary optimization is
\begin{equation}\label{eq:ao_rlr}
  \begin{aligned}
  \phi(g,h)&=\min_{e\in\R^n}\max_{u\in\R^m}
  \Bigl\{\frac{\norm{e}_2}{\sqrt{m}}g^\top u
  -\frac{\norm{u}_2}{\sqrt{m}}h^\top e\\
  &\hspace{2.4em}
  +u^\top w-\frac12\norm{u}_2^2+\lambda F(x_0+e)\Bigr\},
  \end{aligned}
\end{equation}
where $g\sim\mathcal{N}(0,I_m)$ and
$h\sim\mathcal{N}(0,I_n)$ are independent\footnote{We
write the negative sign on the $h$-term. Since all Gaussian variables
are symmetric, this is distributionally equivalent to
\eqref{eq:po}--\eqref{eq:aux}.}.
The corresponding limiting deterministic scalar saddle is,
\begin{equation}\label{eq:D_tau_beta}
  \begin{aligned}
  \max_{\beta\ge 0}\min_{\tau>0}\quad
  &\frac{\delta\beta\sigma_w^2}{2\tau}
    +\frac{\delta\beta\tau}{2}
    -\frac{\delta\beta^2}{2}
    -\frac{\beta\tau}{2}
    +\mathcal{M}(\tau,\beta),
  \end{aligned}
\end{equation}
where the envelope term is
\begin{equation}\label{eq:M_def}
  \begin{aligned}
  \mathcal{M}(\tau,\beta)
  &:=\E_{X_0,H}\!\Bigl[
      \min_{x\in\R}\left\{
      \frac{\beta}{2\tau}(x-X_0+\tau H)^2
      +\lambda f(x)\right\}\Bigr].
  \end{aligned}
\end{equation}
Here, $X_0\sim p_{X_0}$ and
$H\sim\mathcal{N}(0,1)$ is independent of $X_0$.
Let $(\tau_*,\beta_*)$ denote the saddle point and define
\begin{equation}\label{eq:theta_star}
  \theta_*:=\frac{\lambda\tau_*}{\beta_*},
  \qquad
  \sigma_*^2:=\tau_*^2-\sigma_w^2.
\end{equation}
The inner minimization in the envelope term \eqref{eq:M_def} is
equivalent to the proximal problem defined in \eqref{eq:prox}. For fixed
$(\tau,\beta)$, the minimizer in \eqref{eq:M_def} is
$\eta(X_0-\tau H;\lambda\tau/\beta)$. The limits $(\beta_*,\sigma_*,\tau_*)$ are the normalized AO norms
of $(u,e,w+\|e\|_2g/\sqrt m)$, respectively (see
\eqref{eq:amp_finite_coefficient_limits}).

\begin{lemma}[\cite{thrampoulidis2018mestimators}]
\label{lem:cgmt_stationarity}
Suppose $(\tau_*,\beta_*)\in(0,\infty)^2$ is a saddle point of
\eqref{eq:D_tau_beta} and the scalar objective is differentiable
there.  Then the CGMT scalar equations
satisfy
\begin{equation}\label{eq:cgmt_se}
  \tau_*^2=\sigma_w^2+\frac{1}{\delta}
  \E\!\left[\bigl(\eta(X_0-\tau_*H;\theta_*)-X_0\bigr)^2\right],
\end{equation}
and
\begin{equation}\label{eq:cgmt_onsager}
  \frac{\beta_*}{\tau_*}
  =
  1-\frac{1}{\delta}
  \E\!\left[\eta'(X_0-\tau_*H;\theta_*)\right].
\end{equation}
The proof is given in Appendix~\ref{app:proofs} for completeness.
\end{lemma}

\section{Main Results}
\label{sec:main}

In this section, we present our main results.
Theorem \ref{thm:main_draft} states that at the scalar CGMT saddle point,
the AO and PO optimality conditions together have the structure of an
AMP fixed point.

We first state the assumptions under which the CGMT PO and AO
can be compared in terms of their optimizers.  The motivation is the high-dimensional interpretation of the
CGMT: for Gaussian $A$, the PO and AO optimal values and optimizer
properties concentrate around the same deterministic scalar saddle.
Thus, in the proportional limit, it is natural to study a 
setting in which there exist realizations of the auxiliary Gaussian
vectors $g,h$ for which the AO and PO select the same primal-dual
solution.

\begin{assumption}
\label{asm:ao_po_matching}
Let $(e_{\rm AO}^*,u_{\rm AO}^*)$ and
$(e_{\rm PO}^*,u_{\rm PO}^*)$ be selected, not necessarily unique,
saddle points of the AO
\eqref{eq:ao_rlr} and PO \eqref{eq:po_rlr}, respectively.  There exist
specific auxiliary vectors $g,h$ in \eqref{eq:ao_rlr} such that the
selected KKT points match exactly:
\[
 e_{\rm AO}^*=e_{\rm PO}^*=:e^*,
 \qquad
 u_{\rm AO}^*=u_{\rm PO}^*=:u^*.
\]
\end{assumption}

After Assumption~\ref{asm:ao_po_matching}, the remaining issue is
that the regularizer may be nonsmooth.  Although the matched AO and PO
optimizers have the same $x^*:=x_0+e^*$, their stationarity equations
may select different elements of $\partial F(x^*)$.  We therefore
assume that they select a common subgradient.  This is automatic when
$F$ is differentiable at $x^*$.
\begin{assumption}
\label{asm:kkt_selection}
Under the matched optimizers in
Assumption~\ref{asm:ao_po_matching}, the AO and PO primal stationarity
equations select the same regularizer subgradient at the common
optimizer.  That is, for $x^*:=x_0+e^*$, there exists
$s^*\in\partial F(x^*)$ that satisfies both stationarity conditions,
written explicitly in \eqref{eq:ao_kkt_for_amp} and
\eqref{eq:po_kkt_for_amp}.
\end{assumption}

Assumption~\ref{asm:ao_po_matching} implicitly couples $g$ and
$h$.  We nevertheless assume that they remain independent standard
Gaussian vectors and that $(g,h)$ is independent of $(x_0,w)$.  A
possible route to removing both assumptions is discussed after
Theorem~\ref{thm:vector_gaussian_draft}.

\begin{theorem}
\label{thm:main_draft}
Assume the setup of Section~\ref{sec:setup}, and suppose that
\eqref{eq:D_tau_beta} has a unique interior saddle point
$(\tau_*,\beta_*)$ with $\sigma_*^2=\tau_*^2-\sigma_w^2>0$.
Suppose Assumptions~\ref{asm:ao_po_matching}
and \ref{asm:kkt_selection} hold, and set $x^*:=x_0+e^*$.
Then, with high probability, there exist a scalar
$\widehat\mu>0$ and vectors $v^*,z^*$ satisfying the fixed-point system
\begin{align}
  v^* &= x^*+A^\top z^*, \label{eq:main_fp_v}\\
  x^* &= \eta(v^*;\lambda/\widehat\mu), \label{eq:main_fp_x}\\
  z^* &= y-Ax^*
	    +(1-\widehat\mu)z^*.
	    \label{eq:main_fp_z}
\end{align}
Moreover, as $m,n\to\infty$, $\widehat\mu\xrightarrow{p}\beta_*/\tau_*$
and, if $(v,\theta)\mapsto\eta'(v;\theta)$ is continuous at
$(X_0+\tau_*H,\theta_*)$ almost surely,
\begin{equation}\label{eq:main_mu_amp_limit}
  1-\widehat\mu
  -
  \frac{1}{\delta_n}
  \bigl\langle\eta'(v^*;\lambda/\widehat\mu)\bigr\rangle
  \xrightarrow{p}0 .
\end{equation}
Consequently, successive substitution of \eqref{eq:main_fp_v}--\eqref{eq:main_fp_z} gives the AMP recursion \eqref{eq:ampv1}--\eqref{eq:ampz1} asymptotically.
\hfill$\square$
\end{theorem}

We next identify the AO
vectors $g,h$ with the AMP channel noise. The proof is given in
Appendix~\ref{app:proofs}.

\begin{theorem}
\label{thm:vector_gaussian_draft}
Assume the setup of Theorem~\ref{thm:main_draft}, and let
$(v^*,z^*)$ be the fixed-point variables in
\eqref{eq:main_fp_v}--\eqref{eq:main_fp_z}.  Then, with high probability,
there exist scalars $\widehat\tau$, $\widehat\sigma$, and
$\widehat k$ such that the selected AO Gaussian vectors $g,h$ satisfy
\begin{align}
  v^* &= x_0+\widehat\tau h, \label{eq:kkt_h_identity}\\
	  \widehat k z^*
  &= w+\widehat\sigma g, \label{eq:kkt_g_identity}
\end{align}
where, as $m,n\to\infty$,
$\widehat\tau\xrightarrow{p}\tau_*$ and
$\widehat\sigma\xrightarrow{p}\sigma_*$, while
$\widehat k\xrightarrow{p}1$.
\hfill$\square$
\end{theorem}

Motivated by the CGMT-based characterization of the debiased
LASSO in \cite{miolane2021distribution}, a possible route to recover the AMP
fixed-point relations in \eqref{eq:main_fp_z} without Assumptions~\ref{asm:ao_po_matching}
and \ref{asm:kkt_selection} is to set
$z_{\rm PO}^*:=(\tau_*/\beta_*)(y-A\hat x)$ and
$v_{\rm PO}^*:=\hat x+A^\top z_{\rm PO}^*$.
The PO KKT \eqref{eq:po_kkt_for_amp} already gives
$\hat x=\eta(v_{\rm PO}^*;\theta_*)$ directly.
The remaining step is to prove that the empirical distribution of
$((v_{\rm PO}^*)_i,x_{0,i})$ converges to that of
$(X_0+\tau_*H,X_0)$, with $H$ independent of $X_0$.
Under the continuity condition in Theorem~\ref{thm:main_draft},
this convergence and \eqref{eq:cgmt_onsager} would then give
$1-\beta_*/\tau_*$ as the limit of
$\delta_n^{-1}\langle\eta'(v_{\rm PO}^*;\theta_*)\rangle$,
thus asymptotically recovering \eqref{eq:main_fp_z}.

\section{Extension to General Output Losses}
\label{sec:general_extension}

We now extend the analysis to non-quadratic output losses. For a separable, proper, closed and convex loss $L(r)=\sum_{i=1}^m\ell(r_i)$, the estimator of interest is
\begin{equation}\label{eq:smooth_loss_problem}
  \hat{x} \in \arg\min_{x\in\R^n} L(y-Ax)+\lambda F(x).
\end{equation}
Define the scalar maps
\[
  g_{\rm out}(a,\tau_p)
  :=\frac{a-\prox_{\tau_p\ell}(a)}{\tau_p},
  \qquad
  g_{\rm in}(r,\tau_r)
  :=\prox_{\lambda\tau_r f}(r),
\]
which act coordinatewise. The PO--AO construction, scalar
conditions, and proofs are given in Appendix~\ref{app:general_loss}.
\begin{theorem}
\label{thm:gamp_extension}
Suppose Assumptions~\ref{asm:ao_po_matching_gen}--
\ref{asm:gamp_scalar_regularity} hold, and set $x^*:=x_0+e^*$.
Then, with high probability, there exist positive scalars
$\widehat\kappa,\widehat\mu_r$ and vectors $a^*,r_{\rm in}^*$
such that the PO and AO optimality conditions yield the fixed-point system
\begin{align}
  a^*&=y-Ax^*+\widehat\kappa u^*, \label{eq:gamp_a_main}\\
  u^*&=g_{\rm out}(a^*,\widehat\kappa), \label{eq:gamp_u_main}\\
  r_{\rm in}^*&=x^*+\widehat\mu_r A^\top u^*, \label{eq:gamp_r_main}\\
  x^*&=
  g_{\rm in}(r_{\rm in}^*,\widehat\mu_r).
  \label{eq:gamp_x_main}
\end{align}
Moreover, if the almost-sure joint continuity conditions in
Lemma~\ref{lem:gamp_onsager_identities}(ii) hold, then, as $m,n\to\infty$,
\begin{align}
  \frac{1}{\widehat\mu_r}
  -\bigl\langle
  \partial_a g_{\rm out}(a^*,\widehat\kappa)
  \bigr\rangle
  &\xrightarrow{p}0,\label{eq:gamp_tau_r_main}\\
  \widehat\kappa
  -\frac{\widehat\mu_r}{\delta_n}
  \bigl\langle
  \partial_{r_{\rm in}}g_{\rm in}(r_{\rm in}^*,\widehat\mu_r)
  \bigr\rangle
  &\xrightarrow{p}0.
  \label{eq:gamp_tau_p_main}
\end{align}
These equations have the fixed-point form of scalar-variance
max-sum GAMP
\cite[Alg.~2, Eqs.~(9b),(10a),(11b),(12a)]{rangan2011generalized},
with the Onsager relations recovered asymptotically and scalar
functions as in
\cite[Eqs.~(18),(24)--(26)]{rangan2011generalized}. \hfill$\square$
\end{theorem}

\begin{theorem}
\label{thm:gamp_channel_extension}
Assume the setup of Theorem~\ref{thm:gamp_extension}, and let
$(\sigma_*,\tau_{g,*},\beta_*,\tau_{h,*})$ be the saddle point of
\eqref{eq:gen_scalar_ao_final}. The scalars and vectors in
Theorem~\ref{thm:gamp_extension} can be chosen so that, with high
probability, there exist scalars $\widehat\tau,\widehat\sigma$
satisfying
\begin{align}
  r_{\rm in}^*&=x_0+\widehat\tau h,
  \label{eq:gamp_h_identity}\\
  a^*&=w+\widehat\sigma g,
  \label{eq:gamp_g_identity}
\end{align}
where, as $m,n\to\infty$,
\begin{equation}\label{eq:gamp_channel_coefficient_limits}
\begin{aligned}
  \widehat\tau&\xrightarrow{p}
  \frac{\sqrt\delta\,\sigma_*\beta_*}{\tau_{h,*}},
  &\qquad
  \widehat\sigma&\xrightarrow{p}\sigma_*.
\end{aligned}
\end{equation}
\end{theorem}

Taken together, Theorem~\ref{thm:main_draft} (and
Theorem~\ref{thm:gamp_extension}) recover the AMP (and GAMP) fixed
points using the CGMT framework for regularized $M$-estimation.
Theorem~\ref{thm:vector_gaussian_draft} and
(Theorem~\ref{thm:gamp_channel_extension}) then provide an
interpretation of AO Gaussian vectors as the channel noise in the AMP
(and GAMP) channels.

\onecolumn
\raggedbottom
\appendices
\colorlet{appendixheadingcolor}{black}
\makeatletter
\renewcommand{\subsection}{\@startsection{subsection}{2}{\z@}
  {3.5ex plus 1.5ex minus 1.5ex}{0.7ex plus .5ex}
  {\normalfont\normalsize\bfseries\color{appendixheadingcolor}}}
\renewcommand{\subsubsection}{\@startsection{subsubsection}{3}{\z@}
  {1.5ex plus .5ex minus .5ex}{0.5ex}
  {\normalfont\normalsize\bfseries\color{appendixheadingcolor}}}
\renewcommand{\paragraph}{\@startsection{paragraph}{4}{\z@}
  {0.7ex plus .3ex minus .2ex}{0ex}
  {\normalfont\normalsize\bfseries\color{appendixheadingcolor}}}
\makeatother
\newcommand{\appheading}[1]{\textcolor{appendixheadingcolor}{\textbf{#1}}}

\section{Proofs for the Squared-Loss Problem}
\label{app:proofs}

We prove Theorems~\ref{thm:main_draft} and
\ref{thm:vector_gaussian_draft} in four stages. First,
Appendix~\ref{subsec:amp_finite} compares the KKT conditions of the PO
\eqref{eq:po_rlr} and AO \eqref{eq:ao_rlr} at the common optimizer and
common regularizer subgradient in Assumptions~\ref{asm:ao_po_matching}
and \ref{asm:kkt_selection}.  This gives, exactly for finite $m,n$, the
AMP fixed-point equations \eqref{eq:main_fp_v}--\eqref{eq:main_fp_z}
and Gaussian-channel identities \eqref{eq:kkt_h_identity}--
\eqref{eq:kkt_g_identity}.  Second,
Appendix~\ref{subsec:amp_scalar} derives the equations
\eqref{eq:cgmt_se}--\eqref{eq:cgmt_onsager} from the scalar AO
\eqref{eq:D_tau_beta}.  Third,
Appendix~\ref{subsec:amp_coefficients} proves the various scalar
limits \eqref{eq:amp_finite_coefficient_limits}.  Finally,
Appendix~\ref{subsec:amp_empirical} uses the Gaussian-channel identity \eqref{eq:kkt_h_identity}--\eqref{eq:kkt_g_identity}
and these limits to prove the empirical Onsager relation
\eqref{eq:main_mu_amp_limit}.
The fixed-point equations
\eqref{eq:main_fp_v}--\eqref{eq:main_fp_z}, coefficient limits
\eqref{eq:amp_finite_coefficient_limits}, and empirical Onsager
relation \eqref{eq:main_mu_amp_limit} prove
Theorem~\ref{thm:main_draft}. The Gaussian-channel identities
\eqref{eq:kkt_h_identity}--\eqref{eq:kkt_g_identity} and coefficient
limits \eqref{eq:amp_finite_coefficient_limits} prove
Theorem~\ref{thm:vector_gaussian_draft}.

\subsection{Comparing PO and AO KKT conditions}
\label{subsec:amp_finite}
\subsubsection[PO stationarity]{PO stationarity}
The PO \eqref{eq:po_rlr} stationarity conditions are
\begin{equation}\label{eq:po_kkt_for_amp}
 u^*=y-Ax^*,\qquad A^\top u^*=\lambda s^*,
 \qquad s^*\in\partial F(x^*).
\end{equation}
\subsubsection[AO stationarity]{AO stationarity}
In the following discussion, we assume $\norm{e^*}_2>0$ and
$g^\top u^*>0$. Appendix~\ref{subsec:amp_coefficients} shows that
these conditions hold with high probability.
\paragraph[Stationarity in e]{Stationarity in $e$}
The AO
\eqref{eq:ao_rlr} $e$-stationarity condition, with the common
subgradient in Assumption~\ref{asm:kkt_selection}, is
\begingroup
\begin{equation}\label{eq:kkt_ao_e_stationarity}
 \underbrace{\frac{g^\top u^*}{\sqrt m\,\norm{e^*}_2}}
 _{=:\widehat\mu}\,e^*
 -\underbrace{\frac{\norm{u^*}_2}{\sqrt m}}
 _{=:\widehat\beta}\,h+\lambda s^*=0.
\end{equation}
\endgroup
Dividing by $\widehat\mu>0$ and setting
\begingroup
\begin{equation}\label{eq:kkt_coefficients_amp}
 \widehat\tau:=\frac{\widehat\beta}{\widehat\mu},
 \qquad
 \widehat\theta:=\frac{\lambda}{\widehat\mu}
\end{equation}
\endgroup
gives
\begin{equation}\label{eq:ao_kkt_for_amp}
 x_0+\widehat\tau h-x^*=\widehat\theta s^*.
\end{equation}
\begingroup
Since $\widehat\mu>0$, \eqref{eq:ao_kkt_for_amp} is the optimality
condition for
\begin{equation}\label{eq:ao_proximal_minimization}
 x^*=\argmin_{x\in\R^n}
 \left\{\frac12
 \norm{x-\left(x_0+\frac{\widehat\beta}{\widehat\mu}h\right)}_2^2
 +\frac{\lambda}{\widehat\mu}F(x)\right\}.
\end{equation}
By separability and \eqref{eq:prox}, the AO therefore identifies
the proximal input and map as
\begin{equation}\label{eq:ao_proximal_input}
 v^*:=x_0+\frac{\widehat\beta}{\widehat\mu}h,
 \qquad
 x^*=\eta\!\left(v^*;\frac{\lambda}{\widehat\mu}\right).
\end{equation}
Comparing \eqref{eq:po_kkt_for_amp} and \eqref{eq:ao_kkt_for_amp}
with the common subgradient gives
\begin{equation}\label{eq:po_ao_subgradient_comparison}
 v^*-x^*=\frac{\lambda}{\widehat\mu} s^*=\frac{1}{\widehat\mu}A^\top u^*.
\end{equation}
Setting $z^*:=u^*/\widehat\mu$ now gives
\endgroup
\begin{equation}\label{eq:main_vz_underbrace}
 v^*=x^*+A^\top z^*=x_0+\widehat\tau h.
\end{equation}
\begingroup
Also, $u^*=\widehat\mu z^*=y-Ax^*$ gives
\begin{equation}\label{eq:po_rescaled_residual}
 z^*=y-Ax^*+(1-\widehat\mu)z^*.
\end{equation}
\endgroup
This proves \eqref{eq:main_fp_v}--\eqref{eq:main_fp_z}
and \eqref{eq:kkt_h_identity} exactly for finite $m,n$. We will show later in Appendix~\ref{subsec:amp_empirical}, using the scalar identity
\eqref{eq:scalar_stein_identity} and the coefficient limits
\eqref{eq:amp_finite_coefficient_limits}, that the coefficient $1-\widehat\mu$ asymptotically equals the AMP Onsager
coefficient $\delta_n^{-1}\langle\eta'(v^*;\widehat\theta)\rangle$.
\paragraph[Stationarity in u]{Stationarity in $u$}
The AO
$u$-stationarity condition is
\begingroup
\begin{equation}\label{eq:kkt_ao_u_simplified}
 w+\underbrace{\frac{\norm{e^*}_2}{\sqrt m}}
 _{=:\widehat\sigma}\,g-u^*-
 \underbrace{\frac{h^\top e^*}{\sqrt m\,\norm{u^*}_2}}
 _{=:\widehat\kappa}\,u^*=0.
\end{equation}
\endgroup
With $u^*=\widehat\mu z^*$ and
$\widehat k:=\widehat\mu(1+\widehat\kappa)$, this reduces to
\begin{equation}
  \widehat k z^*=w+\widehat\sigma g,
\end{equation}
proving \eqref{eq:kkt_g_identity}.

\subsection{Scalar AO and Proof of Lemma~\ref{lem:cgmt_stationarity}}
\label{subsec:amp_scalar}
For completeness, we derive the stationarity equations
\eqref{eq:cgmt_se}--\eqref{eq:cgmt_onsager}, which will be used to
relate the CGMT scalar saddle to the AMP fixed-point equations.

\noindent\appheading{Proof of Lemma~\ref{lem:cgmt_stationarity}.}

\noindent\appheading{Scalar reduction.}
The PO \eqref{eq:po_rlr} and AO \eqref{eq:ao_rlr} have unbounded
domains.  Following
\cite[Eqs.~(66)--(70), Lemmas~5--7]{thrampoulidis2018mestimators}, we
impose the fixed compact bounds required by the CGMT.  Let
$\phi_{K_e,K_u}$ denote the optimal value of the AO
\eqref{eq:ao_rlr} with $\norm{e}_2\le K_e\sqrt m$ and
$\norm{u}_2\le K_u\sqrt m$; this is the compactified AO.  We also
assume, as required in \cite{thrampoulidis2018mestimators}, that
$f(x_-)$ and $f(x_+)$ are finite for some $x_-<0<x_+$.  Lemmas~8--9
of \cite{thrampoulidis2018mestimators} show that sufficiently large
bounds leave the asymptotic optimal costs and the error-norm
conclusion \eqref{eq:po_error_norm_limit} unchanged.
Writing $\norm{u}_2=\beta\sqrt m$ and maximizing over its direction
gives
\begingroup
\begin{equation}\label{eq:ao_directional_maximization}
 \phi_{K_e,K_u}
 =\min_{\norm{e}_2\le K_e\sqrt m}
  \max_{0\le\beta\le K_u}\Biggl\{
  \beta\sqrt m\left\|w+\frac{\norm{e}_2}{\sqrt m}g\right\|_2
  -\beta h^\top e-\frac{m\beta^2}{2}
  +\lambda F(x_0+e)\Biggr\}.
\end{equation}
\endgroup
Writing $\sigma=\norm{e}_2/\sqrt m$, independence and the laws of
large numbers for $\norm{g}_2^2/m$, $g^\top w/m$, and
$\norm{w}_2^2/m$ give, for every finite $K$,
\begin{equation}\label{eq:uniform_gaussian_norm_limit}
 \sup_{0\le\sigma\le K}
 \left|\frac1{\sqrt m}\norm{w+\sigma g}_2
 -\sqrt{\sigma_w^2 + \sigma^2}\right|\xrightarrow{p}0 .
\end{equation}
Introducing $\tau>0$ through
$\sqrt{x}=\min_{\tau>0}\{x/(2\tau)+\tau/2\}$, using the minimax
reordering in \cite[Lemma~8]{thrampoulidis2018mestimators}, and
subtracting the term $\lambda F(x_0)/n$, the
normalized AO objective is
\begingroup
\begin{equation}\label{eq:ao_scalar_variational_form}
  \max_{\beta\ge0}\min_{\tau>0}\min_{e\in\R^n}\Biggl\{
  \frac{\delta_n\beta\sigma_w^2}{2\tau}
  +\frac{\delta_n\beta\tau}{2}
  -\frac{\delta_n\beta^2}{2}
  +\frac1n\sum_{i=1}^n\left[
  \frac{\beta}{2\tau}e_i^2-\beta h_i e_i
  +\lambda\bigl(f(x_{0,i}+e_i)-f(x_{0,i})\bigr)
  \right]\Biggr\}+o_p(1).
\end{equation}
\endgroup
For fixed $(\tau,\beta)$, the minimization over $e$ separates
coordinatewise.  Completing the square gives
\begingroup
\begin{equation}\label{eq:ao_complete_square}
  \frac{\beta}{2\tau}e_i^2-\beta h_i e_i
  +\lambda f(x_{0,i}+e_i)
  =\frac{\beta}{2\tau}(e_i-\tau h_i)^2
  +\lambda f(x_{0,i}+e_i)
  -\frac{\beta\tau}{2}h_i^2 .
\end{equation}
\endgroup
With $x=x_0+e_i$, the minimization for each coordinate is
\begin{equation}\label{eq:ao_coordinate_minimization}
  \min_{x\in\R}\left\{
  \frac{\beta}{2\tau}\bigl(x-(x_0+\tau h)\bigr)^2+\lambda f(x)
  \right\}-\frac{\beta\tau}{2}h^2-\lambda f(x_0).
\end{equation}
Following \cite{thrampoulidis2018mestimators}, center the Moreau envelope
term by defining
\begingroup
\begin{equation}\label{eq:centered_moreau_regularizer}
 \mathcal M_{\rm c}(\tau,\beta)
 :={\E}_{X_0,H}\!\left[
 \min_{x\in\R}\left\{
 \frac{\beta}{2\tau}(x-X_0+\tau H)^2+\lambda f(x)
 \right\}-\lambda f(X_0)\right].
\end{equation}
\endgroup
The subtracted term is independent of $(\tau,\beta)$, so it changes
neither the proximal minimizer nor the stationarity equations and
allows \cite[Lemma~9]{thrampoulidis2018mestimators} to be applied
without separately assuming $\E|f(X_0)|<\infty$.  The main text uses
the equivalent uncentered notation \eqref{eq:M_def}.
Since $h$ and $-h$ have the same distribution, the limiting centered
scalar objective is
\begin{equation}\label{eq:centered_scalar_objective}
  D_{\rm c}(\tau,\beta)
  :=\frac{\delta\beta\sigma_w^2}{2\tau}
    +\frac{\delta\beta\tau}{2}
    -\frac{\delta\beta^2}{2}
    -\frac{\beta\tau}{2}
    +\mathcal{M}_{\rm c}(\tau,\beta).
\end{equation}
By \cite[Lemma~9]{thrampoulidis2018mestimators}, the normalized
finite AO objective converges uniformly to $D_{\rm c}$ on compact
subsets of $(0,\infty)^2$, and its optimization may be restricted to a
sufficiently large compact set.  Thus the limiting AO is
$\max_{\beta\ge0}\min_{\tau>0}D_{\rm c}(\tau,\beta)$, which is
equivalent to \eqref{eq:D_tau_beta} because the centering is independent
of $(\tau,\beta)$.

\noindent\appheading{Stationarity of the scalar objective.}

We now derive \eqref{eq:cgmt_se} and \eqref{eq:cgmt_onsager}
from the stationarity conditions at a positive saddle $(\tau,\beta)$.
Let $\widehat X=\eta(X_0-\tau H;\lambda\tau/\beta)$ and
$V=X_0-\tau H$.  The envelope theorem gives
\begingroup
\begin{equation}\label{eq:centered_ao_tau_stationarity}
  0=\partial_\tau D_{\rm c}
  =-\frac{\delta\beta\sigma_w^2}{2\tau^2}
  +\frac{\delta\beta}{2}-\frac{\beta}{2}
  -\frac{\beta}{2\tau^2}\E[(\widehat X-V)^2]
  +\frac{\beta}{\tau}\E[(\widehat X-V)H].
\end{equation}
\endgroup
Using
\begin{equation}\label{eq:scalar_error_expansion}
  \E[(\widehat X-V)^2]-2\tau\E[(\widehat X-V)H]
  =\E[(\widehat X-X_0)^2]-\tau^2
\end{equation}
yields \eqref{eq:cgmt_se}.

\begingroup
\noindent\textbf{Onsager relation.}
Since $\eta$ is Lipschitz, combining stationarity with Stein's identity gives
\begin{align}
  0&=\partial_\beta D_{\rm c}
  +\frac{\tau}{\beta}\partial_\tau D_{\rm c}
  =\delta(\tau-\beta)+\E[(\widehat X-X_0)H],
  \label{eq:centered_ao_beta_stationarity}\\[0.5ex]
  1-\frac{\beta}{\tau}
  &=-\frac{\E[(\widehat X-X_0)H]}{\delta\tau}
  =\underbrace{\frac1\delta
  \E[\eta'(X_0-\tau H;\lambda\tau/\beta)]}
  _{\text{Onsager coefficient}}.
  \label{eq:scalar_stein_identity}
\end{align}
This is \eqref{eq:cgmt_onsager} at $(\tau_*,\beta_*)$ and
completes the proof of Lemma~\ref{lem:cgmt_stationarity}.
\endgroup
\hfill$\square$

\subsection{Limits of the AO-Selected Coefficients}
\label{subsec:amp_coefficients}

\begingroup
Define
\begin{equation}\label{eq:residual_direction_definitions}
 a^*:=w+\widehat\sigma g,
 \qquad \bar\tau:=\frac{\norm{a^*}_2}{\sqrt m},
\end{equation}
and let
\begin{equation}\label{eq:amp_nondegenerate_event}
 \mathcal C_n:=\{\norm{e^*}_2>0,\ g^\top u^*>0,\ \bar\tau>0\}.
\end{equation}
\endgroup

\begin{lemma}[Finite-coefficient limits]
\label{lem:amp_coefficient_limits}
Under the assumptions of Theorem~\ref{thm:main_draft},
\begingroup
\begin{equation}\label{eq:amp_finite_coefficient_limits}
 \widehat\sigma\xrightarrow{p}\sigma_*,
 \qquad
 \bar\tau,\widehat\tau\xrightarrow{p}\tau_*,
 \qquad
 \widehat\beta\xrightarrow{p}\beta_*,
 \qquad
 \widehat\mu\xrightarrow{p}\frac{\beta_*}{\tau_*},
 \qquad
 \widehat\theta\xrightarrow{p}\theta_*,
 \qquad
 \widehat k\xrightarrow{p}1.
\end{equation}
\endgroup
\end{lemma}

\appheading{Proof.}

\noindent\appheading{Convergence of $\widehat\sigma$ and $\bar\tau$.}

For the optimizer of the PO \eqref{eq:po_rlr},
\cite[Theorem~1]{thrampoulidis2018mestimators} gives
\begin{equation}\label{eq:po_error_norm_limit}
 \widehat\sigma=\frac{\norm{e^*}_2}{\sqrt m}
 \xrightarrow{p}\sigma_*.
\end{equation}
Equation~\eqref{eq:uniform_gaussian_norm_limit} then gives
\begin{equation}\label{eq:bar_tau_limit}
 \bar\tau
 =\frac{\norm{w+\widehat\sigma g}_2}{\sqrt m}
 \xrightarrow{p}\sqrt{\sigma_w^2+\sigma_*^2}=\tau_*.
\end{equation}
\noindent\appheading{Convergence of $\widehat\beta$.}

Note that $\widehat\beta$ is the normalized
norm of $u^*=y-Ax^*$. We identify its limit as follows.
For $\rho>0$, define
\begin{equation}\label{eq:perturbed_po_value}
 V_n(\rho):=\frac1n\min_x\left\{
 \frac{\rho}{2}\norm{y-Ax}_2^2
 +\lambda\bigl(F(x)-F(x_0)\bigr)\right\}.
\end{equation}
\begingroup
Define
\begin{equation}\label{eq:perturbed_scalar_value}
 V(\rho)=\max_{\beta\ge0}\min_{\tau>0}
 \left\{D_{\rm c}(\tau,\beta)
 +\frac{\delta\beta^2}{2}\left(1-\frac1\rho\right)\right\}.
\end{equation}
Using \cite[Remark~10 and Lemmas~8--9]{thrampoulidis2018mestimators},
we have, for fixed $\rho$,
\begin{equation}\label{eq:perturbed_value_convergence}
 V_n(\rho)\xrightarrow{p}V(\rho).
\end{equation}
At $\rho=1$, the last term in \eqref{eq:perturbed_scalar_value}
vanishes, and the scalar problem has the unique saddle
$(\tau_*,\beta_*)$. The envelope theorem gives
\begin{equation}\label{eq:perturbed_value_derivative}
 V'(1)=\left.\frac{\partial}{\partial\rho}
 \left[\frac{\delta\beta_*^2}{2}\left(1-\frac1\rho\right)\right]
 \right|_{\rho=1}
 =\frac{\delta\beta_*^2}{2}.
\end{equation}
Evaluating $V_n(\rho)$ at the optimizer of $V_n(1)$, which is $x^*$, gives
\begin{equation}\label{eq:perturbed_po_comparison}
 V_n(\rho)-V_n(1)
 \le(\rho-1)\frac{\delta_n\widehat\beta^2}{2}.
\end{equation}
Taking $\rho=1\pm\varepsilon$, with $\varepsilon\in(0,1)$, we get,
\begin{equation}
 \frac{V_n(1+\varepsilon)-V_n(1)}{\varepsilon}
 \le\frac{\delta_n\widehat\beta^2}{2}\le
 \frac{V_n(1)-V_n(1-\varepsilon)}{\varepsilon}.
\end{equation}
Letting $n\to\infty$ using \eqref{eq:perturbed_value_convergence}, and then
$\varepsilon\downarrow0$, both bounds approach $\delta\beta_*^2/2$.
Since $\delta_n\to\delta$ and $\widehat\beta\ge0$, we obtain
\endgroup
\begin{equation}\label{eq:beta_hat_limit}
 \widehat\beta\xrightarrow{p}\beta_*.
\end{equation}
\noindent\appheading{Convergence of $\widehat\mu$, $\widehat\tau$, $\widehat\theta$, and $\widehat k$.}
Since $a^*=w+\widehat\sigma g$,
\begin{equation}\label{eq:gta_positive_limit}
 \frac{g^\top a^*}{m}
 =\frac{g^\top w}{m}
  +\widehat\sigma\frac{\norm{g}_2^2}{m}
 \xrightarrow{p}\sigma_*>0.
\end{equation}
\begingroup
Maximizing over the direction of $u$ in the AO
\eqref{eq:ao_rlr} gives
\begin{equation}\label{eq:dual_alignment_identity}
 u^*=\frac{\widehat\beta}{\bar\tau}a^*
 \qquad\text{when }\bar\tau>0.
\end{equation}
Using \eqref{eq:bar_tau_limit} and \eqref{eq:beta_hat_limit},
\endgroup
\begin{equation}\label{eq:gtu_positive_limit}
 \frac{g^\top u^*}{m}
 =\frac{\widehat\beta}{\bar\tau}\frac{g^\top a^*}{m}
 \xrightarrow{p}\frac{\beta_*\sigma_*}{\tau_*}>0.
\end{equation}
Together with \eqref{eq:po_error_norm_limit} and
\eqref{eq:bar_tau_limit}, this proves
$\mathbb P(\mathcal C_n)\to1$.
\begingroup
On $\mathcal C_n$, \eqref{eq:kkt_ao_e_stationarity}--
\eqref{eq:kkt_coefficients_amp} give
\begin{equation}\label{eq:mu_theta_limits}
 \widehat\mu=\frac{g^\top u^*/m}{\widehat\sigma}
 \xrightarrow{p}\frac{\beta_*}{\tau_*},
 \qquad
 \widehat\tau=\frac{\widehat\beta}{\widehat\mu}
 \xrightarrow{p}\tau_*,
 \qquad
 \widehat\theta=\frac{\lambda}{\widehat\mu}
 \xrightarrow{p}\theta_*.
\end{equation}
Equation~\eqref{eq:kkt_ao_u_simplified} also gives
$a^*=(1+\widehat\kappa)u^*$.
Substituting $u^*=(\widehat\beta/\bar\tau)a^*$ yields
\begin{equation}\label{eq:k_hat_limit}
 1+\widehat\kappa=\frac{\bar\tau}{\widehat\beta},
 \qquad
 \widehat k=\widehat\mu(1+\widehat\kappa)
 =\frac{\widehat\mu\bar\tau}{\widehat\beta}
 \xrightarrow{p}1.
\end{equation}
\hfill$\square$
\endgroup

\subsection[Empirical Onsager Coefficient]{Empirical Onsager Coefficient}
\label{subsec:amp_empirical}

We first state a lemma on empirical averages with random parameters.
We then combine it with the coefficient limits in
Appendix~\ref{subsec:amp_coefficients} to prove the convergence in
\eqref{eq:main_mu_amp_limit}.

\begin{lemma}[Random-parameter empirical channels]
\label{lem:empirical_channel}
Let $(T_i,Z_i)$ be i.i.d., with $\mathbb E|Z_1|<\infty$, and suppose
$(\widehat a_N,\widehat b_N)\xrightarrow{p}(a,b)$.  Let
$\nu_{a,b}$ denote the distribution of $(T_1+aZ_1,b)$.  Then
\begin{equation}\label{eq:empirical_channel_weak_limit}
 \frac1N\sum_{i=1}^N
 \delta_{(T_i+\widehat a_N Z_i,\widehat b_N)}
 \Rightarrow \nu_{a,b}
\end{equation}
weakly in probability.  Hence, by the Portmanteau theorem \cite{billingsley1999},
the corresponding empirical averages converge in probability for every
bounded function whose discontinuity set has zero $\nu_{a,b}$-measure.
\end{lemma}

\appheading{Proof.}
With $(a,b)$ fixed, the empirical measure converges almost surely by the
law of large numbers.
To compare the empirical measures with estimated and fixed
parameters, let $\varphi$ be any bounded $1$-Lipschitz function. Then
\begingroup
\begin{equation}\label{eq:empirical_channel_parameter_bound}
 \left|\frac1N\sum_{i=1}^N
 \bigl[\varphi(T_i+\widehat a_N Z_i,\widehat b_N)
 -\varphi(T_i+aZ_i,b)\bigr]\right|
 \le |\widehat a_N-a|\frac1N\sum_{i=1}^N|Z_i|
 +|\widehat b_N-b|\xrightarrow{p}0.
\end{equation}
\endgroup
The right-hand side tends to zero because
$N^{-1}\sum_{i=1}^N|Z_i|\to\mathbb E|Z_1|$ almost surely and the
estimated parameters converge in probability. Since the bound is
uniform in $\varphi$, the empirical measure with estimated parameters
has the same weak limit, in probability.
\hfill$\square$

\begingroup
\noindent\appheading{Proof of \eqref{eq:main_mu_amp_limit}.}
Apply Lemma~\ref{lem:empirical_channel} with
$(T,Z,\widehat a_N,\widehat b_N)
=(X_0,H,\widehat\tau,\widehat\theta)$, using the limits in
\eqref{eq:mu_theta_limits}. By \eqref{eq:kkt_h_identity},
$v^*=x_0+\widehat\tau h$.
For \eqref{eq:main_mu_amp_limit}, Theorem~\ref{thm:main_draft} assumes
that $\eta'(v;\theta)$ is jointly continuous in $(v,\theta)$ at
$(X_0+\tau_*H,\theta_*)$ with probability one as
required by Lemma~\ref{lem:empirical_channel}. Since $0\le\eta'\le1$
and $\delta_n\to\delta$, the lemma gives
\begin{equation}\label{eq:amp_empirical_onsager_average}
 \frac1{\delta_n}\bigl\langle\eta'(v^*;\widehat\theta)\bigr\rangle
 \xrightarrow{p}
 \frac1\delta\mathbb E[\eta'(X_0+\tau_*H;\theta_*)]
 =1-\frac{\beta_*}{\tau_*}.
\end{equation}
The equality follows from \eqref{eq:cgmt_onsager} and Gaussian symmetry.
Combining this with $\widehat\mu\xrightarrow{p}\beta_*/\tau_*$ from
\eqref{eq:mu_theta_limits} proves
\eqref{eq:main_mu_amp_limit}. \hfill$\square$
\endgroup

\colorlet{appendixheadingcolor}{black}
\section{Proofs for General Output Losses}
\label{app:general_loss}

We prove Theorems~\ref{thm:gamp_extension} and
\ref{thm:gamp_channel_extension} in four stages. First,
Appendix~\ref{subsec:gamp_setup} constructs the PO and AO and states
the matching solutions and subgradient assumptions. Second,
Appendix~\ref{subsec:finite_kkt_comparison} compares their KKT
conditions to obtain, exactly for finite $m,n$, the GAMP fixed-point equations
\eqref{eq:gamp_a_main}--\eqref{eq:gamp_x_main} and Gaussian-channel
identities \eqref{eq:gamp_h_identity}--\eqref{eq:gamp_g_identity}. Third,
Appendix~\ref{subsec:gamp_scalar_limits} introduces the scalar AO
\eqref{eq:gen_scalar_ao_final} and proves the coefficient
limits \eqref{eq:gamp_joint_scalar_limits} and
\eqref{eq:gamp_induced_scalar_limits}. Finally,
Appendix~\ref{subsec:gamp_onsager_identities} derives the 
Onsager identities \eqref{eq:gamp_population_output_onsager}--
\eqref{eq:gamp_population_input_onsager} and uses the scalar limits to prove
\eqref{eq:gamp_tau_r_main}--\eqref{eq:gamp_tau_p_main}.
The fixed-point equations
\eqref{eq:gamp_a_main}--\eqref{eq:gamp_x_main} and Onsager relations
\eqref{eq:gamp_tau_r_main}--\eqref{eq:gamp_tau_p_main} prove
Theorem~\ref{thm:gamp_extension}. The Gaussian-channel identities
\eqref{eq:gamp_h_identity}--\eqref{eq:gamp_g_identity} and coefficient
limits \eqref{eq:gamp_joint_scalar_limits} and
\eqref{eq:gamp_induced_scalar_limits} give
\eqref{eq:gamp_channel_coefficient_limits}, proving
Theorem~\ref{thm:gamp_channel_extension}.

\subsection[PO--AO Formulation and Assumptions]{PO--AO Formulation and Assumptions}
\label{subsec:gamp_setup}

This appendix extends the results to the $M$-estimation problem:
\begin{equation}\label{eq:gen_problem}
  \hat{x} \in \arg\min_{x\in\R^n} L(y-Ax)+\lambda F(x) ,
\end{equation}
where $L(v)=\sum_{i=1}^m \ell(v_i)$ is a separable convex loss and
$F(x)=\sum_{j=1}^n f(x_j)$ is a separable convex regularizer.  Let
$e=x-x_0$ and $w=y-Ax_0$.
Following \cite[Sec.~6.3]{thrampoulidis2016mestimators}, introduce
the residual variable $r=w-Ae$ and a multiplier $u$ for this constraint.
With $G=\sqrt m A$, the primary optimization (PO) is
\begingroup
\begin{equation}\label{eq:po_gen_residual}
  \Phi(G)
  =\min_{e\in\R^n,\,r\in\R^m}\max_{u\in\R^m}
  \Bigl\{-\frac{1}{\sqrt m}u^\top G e+u^\top w-u^\top r
  +L(r)+\lambda F(x_0+e)\Bigr\},
\end{equation}
\endgroup
and the corresponding auxiliary optimization (AO) is
\begingroup
\begin{equation}\label{eq:ao_gen_residual}
  \phi(g,h)
  =\min_{e\in\R^n,\,r\in\R^m}\max_{u\in\R^m}
  \Biggl\{\frac{\norm{e}_2}{\sqrt m}g^\top u
  -\frac{\norm{u}_2}{\sqrt m}h^\top e
  +u^\top w-u^\top r+L(r)+\lambda F(x_0+e)\Biggr\}.
\end{equation}
\endgroup

\begin{assumption}[General-output-loss AO--PO matching]
\label{asm:ao_po_matching_gen}
Let $(e_{\rm AO}^*,u_{\rm AO}^*,r_{\rm AO}^*)$ and
$(e_{\rm PO}^*,u_{\rm PO}^*,r_{\rm PO}^*)$ be KKT solutions of the
residual-augmented AO \eqref{eq:ao_gen_residual} and PO
\eqref{eq:po_gen_residual}, respectively.  There exist specific
auxiliary vectors $g,h$ such that these solutions match exactly:
\begingroup
\begin{equation}\label{eq:gamp_matching_saddles}
  e_{\rm AO}^*=e_{\rm PO}^*=:e^*,
  \qquad
  u_{\rm AO}^*=u_{\rm PO}^*=:u^*,
  \qquad
  r_{\rm AO}^*=r_{\rm PO}^*=:r^* .
\end{equation}
\endgroup
\end{assumption}
As in Assumption~\ref{asm:ao_po_matching}, we assume that the
selected $g,h$ are independent standard
Gaussian vectors and are independent of $(x_0,w)$.

\begin{assumption}[General-output-loss subgradient selection]
\label{asm:kkt_selection_gen}
At the matched solution, the PO \eqref{eq:po_gen_residual} and
AO \eqref{eq:ao_gen_residual} $e$-stationarity equations select the
same $s^*\in\partial F(x^*)$, where $x^*:=x_0+e^*$.
\end{assumption}

\subsection[Comparing PO and AO KKT Conditions]{Comparing PO and AO KKT Conditions}
\label{subsec:finite_kkt_comparison}
Under Assumptions~\ref{asm:ao_po_matching_gen} and
\ref{asm:kkt_selection_gen}, we compare the PO and AO stationarity
conditions to obtain the GAMP fixed-point equations
\eqref{eq:gamp_a_main}--\eqref{eq:gamp_x_main} and Gaussian-channel
identities \eqref{eq:gamp_h_identity}--\eqref{eq:gamp_g_identity}.
The following derivation holds under the positivity conditions in
Assumption~\ref{asm:gamp_scalar_regularity}(ii). The coefficients may be
defined arbitrarily when these conditions fail.

\subsubsection[PO stationarity]{PO stationarity}
The $e$, $u$, and $r$ stationarity conditions for the PO
\eqref{eq:po_gen_residual}, with the common
$s^*\in\partial F(x^*)$, are
\begin{align}
  -A^\top u^*+\lambda s^* &=0, \label{eq:po_e_stat_gen}\\
  -Ae^*+w-r^* &=0, \label{eq:po_u_stat_gen}\\
  u^*&\in\partial L(r^*). \label{eq:po_r_stat_gen}
\end{align}

\begingroup
\begin{samepage}
\subsubsection[AO stationarity]{AO stationarity}
\paragraph[Stationarity in e]{Stationarity in $e$}
The AO \eqref{eq:ao_gen_residual} $e$-stationarity condition is
\begin{equation}\label{eq:ao_e_stat_gen}
 \underbrace{\frac{g^\top u^*}{\sqrt m\,\norm{e^*}_2}}
 _{=:\widehat\mu}\,e^*
 -\underbrace{\frac{\norm{u^*}_2}{\sqrt m}}
 _{=:\widehat\beta}\,h+\lambda s^*=0.
\end{equation}
\end{samepage}
Define
\begin{equation}\label{eq:gamp_input_center}
 r_{\rm in}^*:=x_0+\frac{\widehat\beta}{\widehat\mu}h.
\end{equation}
Using $e^*=x^*-x_0$ and \eqref{eq:gamp_input_center}, \eqref{eq:ao_e_stat_gen} can be re-written as,
\begin{equation}\label{eq:gamp_input_proximal_stationarity}
 0\in x^*-r_{\rm in}^*
 +\frac{\lambda}{\widehat\mu}\partial F(x^*).
\end{equation}
Since $\widehat\mu>0$, this is the optimality condition for the proximal map
\begin{equation}\label{eq:f_prox_gen}
 x^*=\prox_{\frac{\lambda}{\widehat\mu}F}(r_{\rm in}^*)
 =g_{\rm in}(r_{\rm in}^*,1/\widehat\mu),
\end{equation}
where the last equality uses separability. Comparing
\eqref{eq:po_e_stat_gen} and \eqref{eq:gamp_input_proximal_stationarity} with the common
subgradient gives
\begin{equation}\label{eq:gamp_po_ao_e_comparison}
 r_{\rm in}^*=x^*+\frac{\lambda}{\widehat\mu}s^*
 =x^*+\frac{A^\top u^*}{\widehat\mu}.
\end{equation}
Setting
$\widehat\mu_r:=1/\widehat\mu$ gives
\eqref{eq:gamp_r_main}--\eqref{eq:gamp_x_main}.

\paragraph[Stationarity in u and r]{Stationarity in $u$ and $r$}
The AO $u$-stationarity condition is
\begin{equation}\label{eq:ao_u_stat_gen}
 w+\underbrace{\frac{\norm{e^*}_2}{\sqrt m}}
 _{=:\widehat\sigma}\,g-r^*
 -\underbrace{\frac{h^\top e^*}{\sqrt m\,\norm{u^*}_2}}
 _{=:\widehat\kappa}\,u^*=0.
\end{equation}
Define
\begin{equation}\label{eq:gamp_output_center}
 a^*:=w+\widehat\sigma g=r^*+\widehat\kappa u^*,
\end{equation}
where the second equality follows from \eqref{eq:ao_u_stat_gen}.
Combining \eqref{eq:gamp_output_center} with the AO $r$-stationarity
condition $u^*\in\partial L(r^*)$ gives
\begin{equation}\label{eq:gamp_output_proximal_stationarity}
 0\in r^*-a^*+\widehat\kappa\partial L(r^*).
\end{equation}
Since $\widehat\kappa>0$, this is the optimality condition for
\begin{equation}\label{eq:g_resolvent_gen}
 r^*=\prox_{\widehat\kappa L}(a^*).
\end{equation}
Substituting \eqref{eq:g_resolvent_gen} into
\eqref{eq:gamp_output_center} gives
\begin{equation}\label{eq:g_prox_gen}
 u^*=\frac{a^*-\prox_{\widehat\kappa L}(a^*)}{\widehat\kappa}
 =g_{\rm out}(a^*,\widehat\kappa),
\end{equation}
where the last equality uses separability. The PO condition
\eqref{eq:po_u_stat_gen} gives $r^*=y-Ax^*$; substituting this into
\eqref{eq:gamp_output_center} yields
\begin{equation}\label{eq:gamp_output_fixed_point}
 a^*=y-Ax^*+\widehat\kappa u^*.
\end{equation}
This proves \eqref{eq:gamp_a_main}--\eqref{eq:gamp_u_main}.
The definitions of $r_{\rm in}^*$ and $a^*$ also give
\begin{equation}\label{eq:gamp_exact_gaussian_channels}
 r_{\rm in}^*=x_0+\frac{\widehat\beta}{\widehat\mu}h,
 \qquad
 a^*=w+\widehat\sigma g.
\end{equation}
This proves \eqref{eq:gamp_h_identity}--\eqref{eq:gamp_g_identity}
with $\widehat\tau:=\widehat\beta/\widehat\mu$.
Thus, the fixed-point equations are
\begin{equation}\label{eq:gamp_fixed_point_system}
\begin{aligned}
 a^*&=y-Ax^*+\widehat\kappa u^*,\\
 u^*&=g_{\rm out}(a^*,\widehat\kappa),\\
 r_{\rm in}^*&=x^*+\widehat\mu_r A^\top u^*,\\
 x^*&=g_{\rm in}(r_{\rm in}^*,\widehat\mu_r).
\end{aligned}
\end{equation}
\endgroup
The coefficients $\widehat\kappa$ and $\widehat\mu_r$ asymptotically
satisfy the GAMP Onsager relations
\eqref{eq:gamp_tau_r_main}--\eqref{eq:gamp_tau_p_main}, as proved in
Appendix~\ref{subsec:gamp_onsager_identities} using the scalar identities
\eqref{eq:gamp_population_output_onsager}--\eqref{eq:gamp_population_input_onsager}
and the coefficient limits \eqref{eq:gamp_induced_scalar_limits}.

\subsection{Scalar Objective and Coefficient Limits}
\label{subsec:gamp_scalar_limits}

We first define the scalar AO objective and state the assumptions,
including the regularity conditions from
\cite[Eqs.~(7)--(12)]{thrampoulidis2018mestimators}.
We then prove the coefficient limits in
\eqref{eq:gamp_induced_scalar_limits} by showing that the AO scalars
form a saddle of $R_n$ in \eqref{eq:gamp_finite_scalar_objective}
and converge to $p_*$ in \eqref{eq:gamp_positive_scalar_saddle}.

Let $G,H\sim\mathcal N(0,1)$ be independent of $(W,X_0)$ and of each
other, and for a scalar convex function $\varphi$ write
\begin{equation}\label{eq:subgradient_magnitude}
 \varphi'_+(v):=\sup_{s\in\partial\varphi(v)}|s|.
\end{equation}
We use the additive normalizations $\ell(0)=f(0)=0$.
For a proper closed convex scalar function $\varphi$, write
\begin{equation}\label{eq:moreau_envelope_definition}
 e_\varphi(q;t)
 :=\min_v\left\{\frac{(q-v)^2}{2t}+\varphi(v)\right\}.
\end{equation}
Define the centered expected Moreau functions
\begingroup
\begin{equation}\label{eq:centered_moreau_loss_regularizer}
 \mathcal L(c,t)
 :=\mathbb E[e_\ell(cG+W;t)-\ell(W)],
 \qquad
 \mathcal F(c,t)
 :=\mathbb E[e_f(X_0+cH;t)-f(X_0)].
\end{equation}
\endgroup
The deterministic scalar AO is
\begin{equation}\label{eq:gen_scalar_ao_final}
 \inf_{\sigma\ge0,\tau_g>0}
 \sup_{\beta\ge0,\tau_h>0}
 D_{\rm gen}(\sigma,\tau_g,\beta,\tau_h),
\end{equation}
where,
\begingroup
\begin{equation}\label{eq:gen_scalar_D_final}
 D_{\rm gen}
 :=\frac{\delta\beta\tau_g}{2}
 +\delta\mathcal L\left(\sigma,\frac{\tau_g}{\beta}\right)
 -\frac{\sigma\sqrt\delta\,\tau_h}{2}
 -\frac{\sigma\sqrt\delta\,\beta^2}{2\tau_h}
 +\lambda\mathcal F\left(
 \frac{\sigma\sqrt\delta\,\beta}{\tau_h},
 \frac{\sigma\sqrt\delta\,\lambda}{\tau_h}\right),
\end{equation}
\endgroup
\begingroup
with $\mathcal L$ and $\mathcal F$ defined in
\eqref{eq:centered_moreau_loss_regularizer}. Denote a saddle of \eqref{eq:gen_scalar_ao_final} by
\begin{equation}\label{eq:gamp_positive_scalar_saddle}
 p_*=(\sigma_*,\tau_{g,*},\beta_*,\tau_{h,*}).
\end{equation}
\endgroup

\begin{assumption}[Regularity, positivity, and uniqueness]
\label{asm:gamp_scalar_regularity}
\begingroup
\noindent(i) The loss $\ell$ is minimized at zero. In addition to the
finite second moments in Section~\ref{sec:setup}, assume
\begin{equation}\label{eq:gamp_primitive_transfer_conditions}
 \mathbb E\!\left[\ell'_+(W+cG)^2\right]<\infty,
 \qquad
 \mathbb E\!\left[f'_+(X_0+cH)^2\right]<\infty,
 \qquad c\in\mathbb R,
\end{equation}
and that there are $x_-<0<x_+$ for which
$0\le f(x_-),f(x_+)<\infty$.
These are the moment and domain conditions of
\cite[Eqs.~(7)--(12)]{thrampoulidis2018mestimators}.

\noindent(ii) With high probability as $m,n\to\infty$,
$g^\top u^*>0$ and $h^\top e^*>0$.

\noindent(iii) The scalar problem \eqref{eq:gen_scalar_ao_final}
has a unique saddle $p_*$, with strictly positive coordinates.
\endgroup
\end{assumption}

\medskip
\begin{lemma}[Coefficient limits]
\label{lem:gamp_scalar_limits}
Under Assumptions~\ref{asm:ao_po_matching_gen},
\ref{asm:kkt_selection_gen}, and
\ref{asm:gamp_scalar_regularity}, define, based on the saddle point from \eqref{eq:gamp_positive_scalar_saddle},
\begingroup
\begin{equation}\label{eq:gamp_population_coefficients}
 \kappa_*:=\frac{\tau_{g,*}}{\beta_*},
 \qquad
 \mu_{r,*}:=\frac{\sqrt\delta\,\sigma_*}{\tau_{h,*}},
 \qquad
 \tau_{{\rm in},*}:=\mu_{r,*}\beta_*.
\end{equation}
The coefficients $\widehat\kappa$ from \eqref{eq:ao_u_stat_gen},
$\widehat\mu_r=1/\widehat\mu$, and
$\widehat\tau=\widehat\mu_r\widehat\beta$, with
$\widehat\mu,\widehat\beta$ defined in \eqref{eq:ao_e_stat_gen}, satisfy
\begin{equation}\label{eq:gamp_induced_scalar_limits}
  \widehat\kappa\xrightarrow{p}\kappa_*,
  \qquad
  \widehat\mu_r\xrightarrow{p}\mu_{r,*},
  \qquad
  \widehat\tau\xrightarrow{p}\tau_{{\rm in},*}.
\end{equation}
\endgroup
\end{lemma}

\Needspace{12\baselineskip}
\appheading{Proof.}

\appheading{Step 1: empirical scalar objective.}
Define the empirical versions of \eqref{eq:centered_moreau_loss_regularizer} as,
\begingroup
\begin{equation}\label{eq:gamp_empirical_moreau_summaries}
 \mathcal L_n(c,t)
 :=\frac1n\sum_{i=1}^m
 \bigl[e_\ell(cg_i+w_i;t)-\ell(w_i)\bigr],
 \qquad
 \mathcal F_n(c,t)
 :=\frac1n\sum_{j=1}^n
 \bigl[e_f(x_{0,j}+ch_j;t)-f(x_{0,j})\bigr].
\end{equation}
\endgroup
Following \eqref{eq:ao_gen_residual} in
\cite[Appendix~A.6--A.7 and Lemma~8]{thrampoulidis2018mestimators},
we define
\begingroup
\begin{equation}\label{eq:gamp_finite_scalar_objective}
 R_n
 :=\frac{\delta_n\beta\tau_g}{2}
 +\mathcal L_n\left(\sigma,\frac{\tau_g}{\beta}\right)
 -\frac{\sigma\sqrt{\delta_n}\,\tau_h}{2}
 -\frac{\sigma\sqrt{\delta_n}\,\beta^2}{2\tau_h}
   \frac{\norm{h}_2^2}{n}
 +\lambda\mathcal F_n\left(
 \frac{\sigma\sqrt{\delta_n}\,\beta}{\tau_h},
 \frac{\sigma\sqrt{\delta_n}\,\lambda}{\tau_h}\right).
\end{equation}
\endgroup
By \cite[Lemma~8(ii), Lemma~9(a), and the proof of
Lemma~9]{thrampoulidis2018mestimators}, $R_n$ is convex in
$(\sigma,\tau_g)$, concave in $(\beta,\tau_h)$, and, for every compact
$K\subset(0,\infty)^4$,
\begin{equation}\label{eq:gamp_local_uniform_convergence}
 \sup_{p\in K}\left|R_n(p)-D_{\rm gen}(p)\right|
 \xrightarrow{p}0.
\end{equation}
\appheading{Step 2: identification of the saddle.}
\begingroup
We now calculate a saddle of $R_n$ from the AO solution. Define
\begin{equation}\label{eq:kkt_scalars_gen}
 \widehat\tau_g
 :=\frac{\norm{\widehat\sigma g+w-r^*}_2}{\sqrt m}
 =\widehat\kappa\widehat\beta,
 \qquad
 \widehat\tau_h
 :=\frac{\norm{\widehat\beta h-\lambda s^*}_2}{\sqrt n}
 =\widehat\mu\widehat\sigma\sqrt{\delta_n}.
\end{equation}
The equalities follow by taking norms in the AO stationarity
conditions \eqref{eq:ao_e_stat_gen} and \eqref{eq:ao_u_stat_gen}. Thus
\begin{equation}\label{eq:gamp_finite_scalar_ratios}
 \widehat\kappa=\frac{\widehat\tau_g}{\widehat\beta},
 \qquad
 \widehat\mu=\frac{\widehat\tau_h}{\sqrt{\delta_n}\,\widehat\sigma},
 \qquad
 \widehat\mu_r=\frac{\sqrt{\delta_n}\,\widehat\sigma}{\widehat\tau_h}.
\end{equation}
\endgroup

Set
\begin{equation}\label{eq:gamp_empirical_saddle_definition}
 p_n:=(\widehat\sigma,\widehat\tau_g,
       \widehat\beta,\widehat\tau_h).
\end{equation}
We now show that $p_n$ is a saddle of $R_n$. Differentiating
\eqref{eq:gamp_finite_scalar_objective} and using
\eqref{eq:ao_e_stat_gen}, \eqref{eq:ao_u_stat_gen}, and
\eqref{eq:kkt_scalars_gen}--\eqref{eq:gamp_finite_scalar_ratios} gives
\begingroup
\begin{equation}\label{eq:gamp_empirical_saddle_stationarity}
 \begin{aligned}
 \partial_{\tau_g}R_n(p_n)
 &=\frac{\delta_n\widehat\beta}{2}
   -\frac{\norm{u^*}_2^2}{2n\widehat\beta}=0,\\
 \partial_{\tau_h}R_n(p_n)
 &=-\frac{\widehat\sigma\sqrt{\delta_n}}{2}
   +\frac{\norm{e^*}_2^2}
          {2n\widehat\sigma\sqrt{\delta_n}}=0,\\
 \partial_\sigma R_n(p_n)
 &=\frac{g^\top u^*}{n}
   -\sqrt{\delta_n}\widehat\tau_h=0,\\
 \partial_\beta R_n(p_n)
 &=\delta_n\widehat\tau_g-\frac{h^\top e^*}{n}=0.
\end{aligned}
\end{equation}
\endgroup
Thus $p_n = (\widehat\sigma,\widehat\tau_g, \widehat\beta,\widehat\tau_h)$ is a saddle of $R_n$. Write
$q_n=(\widehat\sigma,\widehat\tau_g)$ and
$r_n=(\widehat\beta,\widehat\tau_h)$, so that $p_n=(q_n,r_n)$.
Since $p_n$ is positive, these stationarity equations and the
convexity--concavity of $R_n$ give, for all $q,r\in(0,\infty)^2$,
\begin{equation}\label{eq:gamp_finite_saddle_inequalities}
 R_n(q_n,r)\le R_n(q_n,r_n)\le R_n(q,r_n).
\end{equation}

\appheading{Step 3: convergence to $p_*$.}
It remains to show that the saddle $p_n$ of $R_n$ converges
to the saddle $p_*$ of $D_{\rm gen}$. Equation~\eqref{eq:gamp_local_uniform_convergence}
states that the values of $R_n$ and $D_{\rm gen}$ become uniformly
close on fixed compact sets, whereas $p_n$ is obtained by
optimizing $R_n$ over all positive scalars. We therefore first
restrict $R_n$ to a compact $K\subset(0,\infty)^4$
containing $p_*$ in its interior. We expect this restriction to
leave the limiting solution unchanged. To establish this, we first
prove convergence of the saddles of $R_n$ restricted to $K$ and
then use convexity--concavity to remove the restriction.

Continuity and convexity--concavity ensure that $R_n$ restricted
to $K$ has a saddle for each $n$. Any such saddle
$\widetilde p_n=(\widetilde q_n,\widetilde r_n)$ satisfies, as in
\eqref{eq:gamp_finite_saddle_inequalities},
\begingroup
\begin{equation}\label{eq:gamp_restricted_empirical_saddle_inequalities}
 R_n(\widetilde q_n,r)
 \le R_n(\widetilde q_n,\widetilde r_n)
 \le R_n(q,\widetilde r_n),\qquad (q,r)\in K.
\end{equation}
By \eqref{eq:gamp_local_uniform_convergence}, $R_n$ converges
uniformly in probability to $D_{\rm gen}$ on $K$. The restriction
of $D_{\rm gen}$ to $K$ has the unique saddle
$p_*=(q_*,r_*)$, which satisfies
\begin{equation}\label{eq:gamp_restricted_deterministic_saddle_inequalities}
 D_{\rm gen}(q_*,r)
 \le D_{\rm gen}(q_*,r_*)
 \le D_{\rm gen}(q,r_*),\qquad (q,r)\in K.
\end{equation}
Uniform convergence on the compact set $K$ and uniqueness of
$p_*$ therefore imply that all saddles of $R_n$ restricted to
$K$ converge to $p_*$ in probability:
\begin{equation}\label{eq:gamp_restricted_saddle_convergence}
 \sup_{\substack{p\in K:\\
        p\text{ is a saddle of }R_n|_K}}
 \norm{p-p_*}_2\xrightarrow{p}0.
\end{equation}
\endgroup

With high probability, the saddles of $R_n$
restricted to $K$ lie in its interior and hence are also saddles
on $(0,\infty)^4$. By convexity--concavity, the segment joining
any such saddle to a saddle of $R_n$ outside $K$ would contain
a saddle on the boundary of $K$. Such a saddle remains a positive
distance from $p_*$, contradicting
\eqref{eq:gamp_restricted_saddle_convergence}.
Thus $p_n\xrightarrow{p}p_*$, that is,
\begingroup
\begin{equation}\label{eq:gamp_joint_scalar_limits}
 (\widehat\sigma,\widehat\tau_g,
   \widehat\beta,\widehat\tau_h)
 \xrightarrow{p}
 (\sigma_*,\tau_{g,*},\beta_*,\tau_{h,*}).
\end{equation}
Substituting these limits into \eqref{eq:gamp_finite_scalar_ratios}
gives \eqref{eq:gamp_induced_scalar_limits} since the limiting
denominators are positive.
\endgroup
\hfill$\square$

\subsection[Onsager Coefficients]{Onsager Coefficients}
\label{subsec:gamp_onsager_identities}

We now derive \eqref{eq:gamp_population_output_onsager}--
\eqref{eq:gamp_population_input_onsager} from the stationarity
conditions of $D_{\rm gen}$, and then apply Lemma~\ref{lem:empirical_channel}
to prove \eqref{eq:gamp_tau_r_main}--\eqref{eq:gamp_tau_p_main}.

\begin{samepage}
\begin{lemma}[Onsager identities]
\label{lem:gamp_onsager_identities}
Under Assumptions~\ref{asm:ao_po_matching_gen},
\ref{asm:kkt_selection_gen}, and
\ref{asm:gamp_scalar_regularity}:

(i) We have
\begingroup
\begin{equation}\label{eq:gamp_population_output_onsager}
 \frac1{\mu_{r,*}}
 =
 \mathbb E\!\left[
 \partial_a g_{\rm out}
 (W+\sigma_*G,\kappa_*)\right].
\end{equation}
\begin{equation}\label{eq:gamp_population_input_onsager}
 \kappa_*
 =
 \frac{\mu_{r,*}}{\delta}
 \mathbb E\!\left[
 \partial_r g_{\rm in}
 (X_0+\tau_{{\rm in},*}H,\mu_{r,*})\right].
\end{equation}
\endgroup
(ii) If
$\partial_a g_{\rm out}$ and $\partial_r g_{\rm in}$ are jointly continuous at
\[
 (W+\sigma_*G,\kappa_*),
 \qquad
 (X_0+\tau_{{\rm in},*}H,\mu_{r,*}),
\]
respectively, almost surely, $\widehat\kappa$ and
$\widehat\mu_r$ satisfy
\eqref{eq:gamp_tau_r_main}--\eqref{eq:gamp_tau_p_main}.
\end{lemma}
\end{samepage}

\Needspace{8\baselineskip}
\begingroup
\noindent\appheading{Proof of (i).}
Since $p_*$ is an interior saddle, all partial derivatives of
$D_{\rm gen}$ are zero there. Differentiating
\eqref{eq:gen_scalar_D_final} and combining the $\sigma,\tau_h$
and $\beta,\tau_g$ stationarity conditions, respectively, gives
\begin{equation}\label{eq:gamp_scalar_stationarity_combinations}
 \begin{aligned}
 0&=\partial_\sigma D_{\rm gen}(p_*)
 +\frac{\tau_{h,*}}{\sigma_*}\partial_{\tau_h}D_{\rm gen}(p_*)
 =\delta\mathbb E\!\left[
 Gg_{\rm out}(W+\sigma_*G,\kappa_*)\right]
 -\sqrt\delta\,\tau_{h,*},\\[0.5ex]
 0&=\partial_\beta D_{\rm gen}(p_*)
 +\frac{\tau_{g,*}}{\beta_*}\partial_{\tau_g}D_{\rm gen}(p_*)
 =\delta\tau_{g,*}-\mathbb E\!\left[
 Hg_{\rm in}(X_0+\tau_{{\rm in},*}H,\mu_{r,*})\right],
 \end{aligned}
\end{equation}
where the second identity uses $\mathbb E[HX_0]=0$ and
$\mathbb E[H^2]=1$.
\begingroup
Both scalar maps are Lipschitz for fixed positive parameters.
Using \eqref{eq:gamp_population_coefficients}, $\sigma_*,\beta_*>0$,
and Gaussian integration by parts conditional on $W$ and $X_0$,
respectively, gives the Onsager identities
\eqref{eq:gamp_population_output_onsager}--
\eqref{eq:gamp_population_input_onsager} directly:
\begin{equation}\label{eq:gamp_gaussian_integration_by_parts}
 \begin{aligned}
 \frac1{\mu_{r,*}}
 &=\frac{\mathbb E\!\left[Gg_{\rm out}(W+\sigma_*G,\kappa_*)\right]}{\sigma_*}
 =\mathbb E\!\left[
 \partial_a g_{\rm out}(W+\sigma_*G,\kappa_*)\right],\\[0.5ex]
 \kappa_*
 &=\frac{\mathbb E\!\left[Hg_{\rm in}(X_0+\tau_{{\rm in},*}H,\mu_{r,*})\right]}{\delta\beta_*}
 =\frac{\mu_{r,*}}{\delta}\mathbb E\!\left[
 \partial_r g_{\rm in}(X_0+\tau_{{\rm in},*}H,\mu_{r,*})\right].
 \end{aligned}
\end{equation}
\endgroup
\endgroup

\begingroup
\medskip
\noindent\appheading{Proof of (ii).}
For positive parameters,
\begin{equation}\label{eq:gamp_derivative_bounds}
 0\le\kappa\,\partial_a g_{\rm out}(a,\kappa)\le1,
 \qquad
 0\le\partial_r g_{\rm in}(r,\mu_r)\le1.
\end{equation}
(Set these functions to zero where undefined or the respective
parameter is nonpositive.) Apply Lemma~\ref{lem:empirical_channel}
to these bounded functions with
\begin{equation}\label{eq:gamp_empirical_channel_substitutions}
 (T,Z,\widehat a_N,\widehat b_N)
 =(W,G,\widehat\sigma,\widehat\kappa),
 \qquad
 (T,Z,\widehat a_N,\widehat b_N)
 =(X_0,H,\widehat\tau,\widehat\mu_r),
\end{equation}
using the channel identities \eqref{eq:gamp_exact_gaussian_channels},
the limits \eqref{eq:gamp_joint_scalar_limits} and
\eqref{eq:gamp_induced_scalar_limits}, and the almost-sure joint
continuity assumed in part~(ii). Dividing the
resulting average of $\kappa\,\partial_a g_{\rm out}$ by
$\widehat\kappa\xrightarrow{p}\kappa_*>0$ and using
\eqref{eq:gamp_population_output_onsager}--
\eqref{eq:gamp_population_input_onsager} gives
\begin{equation}\label{eq:gamp_empirical_derivative_limits}
 \left\langle\partial_a g_{\rm out}
 (a^*,\widehat\kappa)\right\rangle
 \xrightarrow{p}\frac1{\mu_{r,*}},
 \qquad
 \left\langle\partial_r g_{\rm in}
 (r_{\rm in}^*,\widehat\mu_r)\right\rangle
 \xrightarrow{p}\frac{\delta\kappa_*}{\mu_{r,*}}.
\end{equation}
Together with \eqref{eq:gamp_induced_scalar_limits} and
$\delta_n\to\delta$, this proves
\eqref{eq:gamp_tau_r_main}--\eqref{eq:gamp_tau_p_main}.
\hfill$\square$
\endgroup

\end{document}